\documentclass[prl, twocolumn]{revtex4}

\usepackage{graphicx}
\usepackage{amsmath}
\usepackage{subfigure}

\def\be{\begin{equation}}
\def\ee{\end{equation}}
\def\bea{\begin{eqnarray}}
\def\eea{\end{eqnarray}}
\newcommand{\eref}[1]{(\ref{#1})}

\newcommand\refeq[1]{Eq.~(\ref{eq:#1})}

\begin{document}

\title{Dependence of resistivity on surface profile in \\nanoscale metal films and wires}
\author{Baruch Feldman, Rui Deng and Scott T. Dunham}
\affiliation{Physics Department, University of Washington, Seattle, WA\\ Electrical Engineering Department, University of Washington, Seattle, WA}

\begin{abstract} 
We extend quantum models of nanowire surface scattering to incorporate bulk resistivity and extract an expression for the increased resistivity due to surface roughness.  To learn how to improve conductivity, we calculate conductivity degradation from individual wavelengths of surface roughness, and show how these can be convolved to give resistivity for arbitrary surfaces.  
We review measurements from Cu films and conclude that roughness at short wavelengths (less than 100 nm) dominates scattering, and that primarily specular scattering should be achievable for RMS roughness below about 0.7 nm.
\end{abstract}

\maketitle

As the minimum feature size in semiconductor technology continues to shrink, metal nanowires with thickness $<$45 nm are now needed to interconnect electronic nanodevices.
However, measurements show nanowires have substantially higher resistivity than bulk metals \cite{Munoz, GrainRefs}, leading to interconnect delays, power loss, and other limits on performance.  
Scattering from surfaces, interfaces and grain boundaries are the causes of this conductivity degradation% in metal wires with dimensions below 100 nm
, but microscopic understanding of these effects and quantitative predictions of their magnitude have been limited.  Here, we investigate the detailed dependence of conductivity on surface roughness profile and analyze the resulting technological impact. 

The first quantitative treatments of surface and size effects in thin films or wires were the semiclassical methods of Fuchs \cite{Fuchs} and Sondheimer \cite{Sondheimer}.  These approaches assume a ratio $p$ of carrier collisions with the surface reflect specularly, while $1-p$ scatter diffusely.  Such theories can be fit to experiment with $p$ as a free parameter, but do not provide insight into how to improve conductivity.  
    
More recently, surface roughness scattering has raised the attention of researchers in industry \cite{ChenGard, Jiang, Kapur}, and quantum mechanical approaches to surface scattering calculations have been proposed.  The two primary approaches include the Kubo linear response theory of Te\u{s}anovi\'{c} \textit{et al.} \cite{Tesanovic} and Trivedi and Ashcroft \cite{Trivedi}, 
and the diagrammatic Keldysh formalism of Meyerovich and collaborators \cite{MeyerStep, MeyerStep-bulk, MeyerPon}.
Here we follow the approach of Meyerovich {\em et al.}, which is readily applied to arbitrary surface roughness profiles.  We calculate the contribution of each spatial frequency of surface roughness and convolve with roughness data extracted from experiments to gain insight into the nature of surface roughness scattering.

%%%%%%%%%%%%%%%%%%%%%%%%%%%%%%%%%%%%%%%%%%%%%%%%%%%%%%%%%%%%%%%%%%%%%%%%%%%%%%%%%%%%%%%%%%%%%%%%%%%

In our conductivity calculations, we consider a thin film because it reproduces the major qualitative results of a wire (and matches quantitatively when Eq.~\eqref{eq:rhoeff-linear} below holds), while avoiding strong localization and other effects that make 1D systems problematic to deal with theoretically \cite{MeyerPon, LeeRama}. For the technologically important 10-100 nm scale, wire conductivity can be accurately estimated by combining effects of scattering from sidewalls to that from top and bottom surfaces.  

In a thin film of thickness $L$, boundary conditions at the surfaces lead to a density of states quantized in the transverse direction.  As a result, the conduction band, described as the set of states at the Fermi energy, is broken into subbands with continuous parallel and quantized transverse components of the Bloch wavevector.  Conduction states are then described by a subband index $j$ and a 2D wavevector $k_j$, subject to the constraint that the total energy is equal to the Fermi energy:
\be
\label{eq:fermi}
E = \frac{\hbar^2}{2 m^*}  \left[\left( \frac{\pi j}{L} \right)^2 + k_j^2\right] = E_F
\ee
(We treat the Fermi surface as effectively spherical, which is particularly appropriate for the best conducting metals, Ag, Cu, and Au).  
Even in a perfectly smooth film, this quantization leads to thickness-dependent conductivity, and to the quantum size effect (QSE), caused by the quantized dependence of the density of states on thickness \cite{Trivedi} which is significant for very thin films ($<$5nm).  

Theoretical approaches to rough surfaces \cite{Tesanovic, MeyerStep} employ a (non-unitary) transformation to map the film with position-dependent surface into a flat film with bulk (non-Hermitian) perturbations.  The scattering depends on the power spectral density (PSD) of the roughness, defined as the Fourier transformed surface height correlation function.  
In \cite{MeyerStep, MeyerPon}, a general isotropic 2D roughness power spectrum $\zeta( | \vec{k} | )$ is treated with diagrammatic perturbation theory.  By Fermi's Golden Rule, the spatial frequencies of roughness determine the interband transition rates and hence a momentum loss rate matrix:
\be
\label{eq:W-def}
W_{j j'}(\chi) = \frac{2 \hbar}{(m^* L)^2} \left( \frac{\pi j}{L} \right)^2 \left( \frac{\pi j'}{L} \right)^2 \; \zeta\left( k_j, k_{j'}, \chi \right),
\ee
\be
\label{eq:tau-def}
\left( \tau_s \right)^{-1}_{jj'} = \frac{m^*}{2} \; \sum_{j''} \left[ \delta_{jj'} W^{(0)}_{j j''} - \delta_{j' j''} W^{(1)}_{j j'} \right].
\ee
Here $\hbar k_j$ is the in-plane momentum satisfying \eref{eq:fermi} for subband $j$, $\chi$ is the angle between initial and final carrier momentum, 
$\zeta\left( k_j, k_{j'}, \chi \right) = \zeta(\sqrt{ k_j^2 + k_{j'}^2 - 2 k_j k_{j'} \cos \chi })$,
and superscripts denote (2D) angular harmonics:
\be
\zeta^{(n)} \equiv \frac{1}{\pi} \int_0^{2 \pi} d\chi \zeta(\chi) \cos(n \chi).
\label{eq:angular-harmonics}
\ee
Surface roughness-limited conductivity is given by
\be
\sigma_s = 1/\rho_s = \frac{\tau_s \: n e^2}{m^*}  = \frac{e^2}{2 \pi \: m^* L} \; \sum_{j j'} k_j \: \tau_{j j'} \: k_{j'},
\label{eq:conductivity}
\ee
where $\rho_s$ is resistivity, the scalar $\tau_s$ is the overall surface relaxation or mean free time, $n=k_F^3 / 3 \pi^2$ is carrier density, and $1/\tau_s \propto \rho_s$ is the overall surface momentum loss rate.  Note that our definition differs from \cite{MeyerPon} by an extra factor of $3/2 \pi L$ because we use the usual 3D conductivity, as in \cite{Trivedi}.  

To combine bulk and surface scattering, we extend the method in \cite{Trivedi}, adding momentum loss rates within subbands, to the case with interband transitions by adding matrices: $\tau^{-1} = \tau^{-1}_b +  \tau^{-1}_s$. Since the primary bulk scattering mechanism at room temperature, acoustic phonons, is nearly isotropic \cite{Tomizawa}, we use 
$$
\left(\tau_b \right)^{-1}_{j j'} = \frac{v_F}{\lambda_b} \delta_{j j'},
$$
with $v_F = 1.6 \times 10^8$ cm/s the Fermi velocity and $\lambda_b$ = 39 nm the bulk mean free path for copper.

Adding matrices produces very different results from adding the scalars $1/\tau \propto \rho$. 
Matthiessen's rule, which states that 
$\rho_T = \rho_1+\rho_2$
for independent (series) resistivity mechanisms, breaks down in thin films when combining bulk and surface scattering~\cite{MeyerStep-bulk, Trivedi, Munoz} (Fig.~\ref{fig:conduct-results} below).  This breakdown can be understood because, absent bulk scattering, conductivity is dominated by carriers with momentum nearly parallel to the surface (low $j$) which rarely scatter from the surface. To consider surface scattering together with bulk scattering, we define 
\be
\rho_s^{\rm eff} \equiv \rho - \rho_b,
\label{eq:cond-result}
\ee
the {\em effective} surface roughness contribution to resistivity, which 
is independent of $\rho_b$ to first order \footnote{
We can show by expanding in $\frac{v_F}{\lambda_b} \tau_s^{-1}$ that $1/\tau_s^{\rm eff} = \frac{2 k_F^3 L}{3 \pi \left( \sum_i k_i^2 \right)^2} \sum_{ij} k_i \left( \tau_s  \right)^{-1}_{ij} k_j \: + \: O\left( \frac{v_F}{\lambda_b} \tau_s^{-1} \right)^2$.}. 

To study the effect of individual spatial frequencies of roughness on resistivity, we perform a first order functional expansion on $\rho_s^{\rm eff}$.
We define the first variation in $\rho_s^{\rm eff}$ with respect to the PSD component at wavevector with magnitude $k_0$  
as the response to a special PSD:
\begin{equation}
\rho_{(\zeta_{k_0})}^{\rm eff} = \frac{l^2}{k_0} \; \frac{\delta \rho^{\rm eff}_s}{\delta \zeta(k_0)}.
\label{eq:fnal-deriv}
\end{equation}
Here $\rho_{(\zeta_{k_0})}^{\rm eff}$ is the response to a 2D PSD of the form
\begin{equation}
\label{eq:delta-PSD}
\zeta_{k_0}({\bf k}) \equiv \frac{l^2 \delta(|{\bf k}| - k_0)}{k_0},
\end{equation}
with $2 \pi l^2$ the mean squared roughness of this PSD. The factor $(l^2/k_0)$ in \eqref{eq:fnal-deriv} is necessary for consistent units.

Consistent with the validity of (\ref{eq:W-def}) -- (\ref{eq:conductivity}) to first order in roughness, we perform a first order functional expansion of $\rho_s^{\rm eff}$ in $\zeta$: 
\begin{equation}
\label{eq:rhoeff-linear}
\rho_{\left(\zeta \right)}^{\rm eff} = \int_0^{\infty} k_0 \; \frac{\rho_{(\zeta_{k_0})}^{\rm eff}}{l^2} \; \zeta(k_0)\: dk_0 \; + \, O \left( \zeta^2 \right)
\end{equation}
Here the LHS is the resistivity from an arbitrary 2D isotropic PSD $\zeta$, and  $\rho_{(\zeta_{k_0})}^{\rm eff}$ in the RHS is given by \refeq{fnal-deriv}. 

The angular harmonics for \eqref{eq:delta-PSD} are given by 
\begin{align*}
\zeta_{k_0}^{(0)}(q, q') &= \frac{2 l^2}{\pi q q' \; | \sin \chi |} \; \theta\left( k_0 - |q-q'| \right) \; \theta(q + q' - k_0), \nonumber \\*
\zeta_{k_0}^{(1)}(q,q') &=  \zeta_{k_0}^{(0)}(q,q') \; \cos\chi, \nonumber
\end{align*}
where $\theta$ is the Heaviside step function, and the delta function sets the angle $\chi$ between the initial and final wavevectors
\[
\cos \chi = \frac{q^2 + q'^2 - k_0^2}{2 q q'}.
\]

Our results for the functional derivative \eqref{eq:fnal-deriv} 
for surface-only scattering $1/\tau_s \propto \rho_s$ and effective rate with bulk scattering $1/\tau_s^{\rm eff} \propto \rho_s^{\rm eff}$ are plotted in Figure \ref{fig:conduct-results}. 
\begin{figure}[b]
  \includegraphics[scale=0.6]{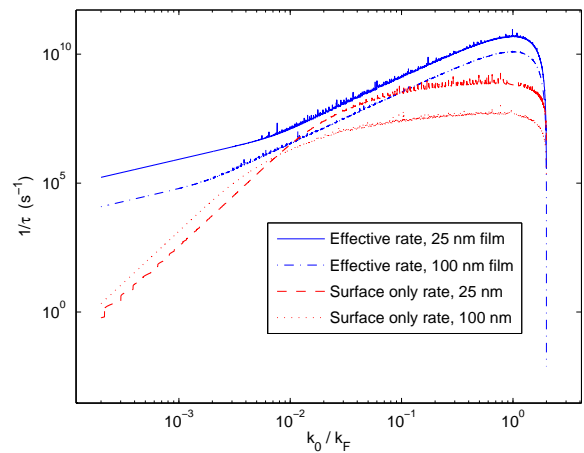}
  \caption{The momentum loss rate $1/\tau_s \propto \rho_s$ and $1/\tau_s^{\rm eff} \propto \rho_s^{\rm eff}$ in response to a single-frequency PSD \eqref{eq:fnal-deriv}, \eqref{eq:delta-PSD} in copper with $l = 10$ pm \:  %of PSDs of the form \eqref{eq:delta-PSD}
  as a function of $k_0$. Note drastic breakdown of Matthiessen's rule, with 
$1/\tau_s^{\rm eff} \gg 1/\tau_s$.
\label{fig:conduct-results} }
\end{figure}
We tested the relation \eqref{eq:rhoeff-linear} for several PSDs, including our fit to experimental surface roughness data (see below).  For the experimental fit, we find  \eqref{eq:rhoeff-linear} holds to 2\% for a 25 nm film and 1\% for 100 nm. We conclude that, for films thick enough that $\rho_s^{\rm eff} \ll \rho$, Eq.~\eqref{eq:rhoeff-linear} provides an effective calculation of resistivities.   

For larger $k_0$ values, which dominate surface scattering, the thickness dependence of the {\em surface only} resistivity is $\rho_{s} \propto L^{-2}$, as in \cite{MeyerPon, Trivedi}.  But for large $L$, we expect $\rho$ to approach the semiclassical result of Fuchs \cite{Fuchs} and Sondheimer \cite{Sondheimer}:
\begin{equation}
\label{eq:FS}
\frac{\rho}{\rho_b}  = \frac{\rho_b + \rho_s^{\rm eff}}{\rho_b} = 1 + \frac{3}{8} \: \frac{\lambda_b}{L} \left( 1 - p \right).
\end{equation}
Fig.~\ref{fig:conduct-results} indeed shows $\rho^{\rm eff}_s \propto k_0^2/L$ for most of the $k_0$ range. 

For very small $k_0$ values, $\rho_s^{\rm eff} \propto k_0 / L^2$, while the surface-only rate actually increases with $L$, $\rho_s \propto L$.  
This can be explained by quantization.  The subbands are spaced closest for lowest $j$, so for $k_0 < \sqrt{3}\pi/(k_F L)$, there is no scattering between higher order subbands.  As $k_0$ becomes smaller, interband scattering due to surface roughness becomes impossible for lower and lower subbands. For $\rho_s$, an increase in $L$ {\em decreases} the spacing between $k_j$ values, introducing interband transitions between low $j$ states where the spacing is smallest.  Physically, this couples states traveling nearly parallel to the film surface with other states that interact much more strongly with the surface, providing a mechanism to increase $\rho_s$ with $L$.  
For $\rho_s^{\rm eff}$, in contrast, electrons in low $j$ states are already frequently scattered by bulk scattering. For small $k_0$, only intraband scattering is possible, so there are always two final states and $\rho_s^{\rm eff}\propto k_0/L^2$.  At higher $k_0$, the number of available final subbands becomes proportional to $L$, so $\rho_s^{\rm eff} \propto k_0^2/L$.

%%%%%%%%%%%%%%%%%%%%%%%%%%%%%%%%%%%%%%%%%%%%%%%%%%%%%%%%%%%%%%%%%%%%%%%%%%%%%%%%%%%%%%%%%%%%%%%%%%%%

\begin{figure}
\includegraphics[scale=0.6]{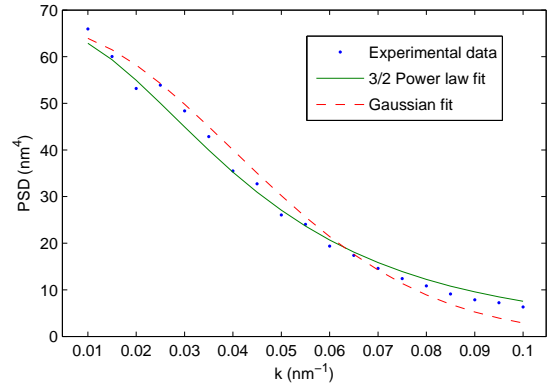}
\caption{Normalized roughness spectra for Cu films %measured by atomic force microscopy (AFM) 
\cite{Moseler} compared to fits to \refeq{feenstra-fit} with correlation length $\Lambda = 18$ nm and Gaussian with $\Lambda = 1/\sigma = 25$ nm. Both forms fit the available data, but give very different predictions of fall-off in roughness at higher spatial frequencies.\label{fig:exp-PSD}} 
\end{figure}

As can be seen from Fig.~\ref{fig:conduct-results}, spatial frequencies near $k_F$ (shortest wavelengths) have the strongest momentum loss, but the impact on conductivity depends on the actual roughness PSD of metal films and wires.  Any attempts to improve conductivity will benefit from a knowledge of which components of surface roughness give the most improvement for the resources spent.  Theoretical calculations often assume a Gaussian roughness spectrum, but experiments show that many PSD forms are present depending on the wire deposition conditions \cite{MeyerPon}, and that real PSDs can fall of more slowly than Gaussian \cite{Feenstra}.  Unfortunately, the experimental literature on surface roughness spectra for metals is limited and focuses on relatively large length scales.  Thus, we look to other materials.  Feenstra {\em et al.} \cite{Feenstra} observed that 1D STM scans of InAs/GaSb superlattice interfaces showed Lorentzian distributions,
$$
\zeta(k) = \frac{2 \Lambda \Delta^2}{\left( 1 + k^2 \Lambda^2 \right)}.
$$
For isotropic roughness, this corresponds to a 2D PSD of the form 
\begin{equation}
\label{eq:feenstra-fit}
\zeta(k) = \frac{2 \pi \Lambda^2 \Delta^2}{\left( 1 + k^2 \Lambda^2 \right)^{3/2}},
\end{equation}
with mean squared roughness $2 \pi \Delta^2$.
Eq.~\eqref{eq:feenstra-fit} also fits the AFM results of Moseler {\em et al.}~\cite{Moseler} for Cu films with $\Lambda=18 \; \text{nm,} \; \Delta = 1.8 \; \text{\AA}$, as shown in Fig.~\ref{fig:exp-PSD}.  Other experiments on copper films confirm a correlation length of $\sim$20 nm~\cite{Keavney02}.  We also fit \cite{Moseler}'s data to a Gaussian PSD, as shown in the figure.  Because measurements at high spatial frequencies (which have a particularly strong effect on scattering) are lacking, our goal is to extrapolate from these fits
\footnote{Ref.~\cite{Feenstra}'s STM of InAs/GaSb, which extends to wavevectors of 10 nm$^{-1}$ compared to \cite{Moseler}'s AFM up to 0.1 nm$^{-1}$, supports the fit in \refeq{feenstra-fit}.  Note that the flattening near the highest frequency is due to the finite sampling interval used in \cite{Moseler}.}.  
%Because of the importance of the roughness at higher wavevectors (Fig.~\ref{fig:conduct-results}), it would be very useful to measure roughness for Cu films/wires on this scale using STM.

%%%%%%%%%%%%%%%%%%%%%%%%%%%%%%%%%%%%%%%%%%%%%%%%%%%%%%%%%%%%%%%%%%%%%%%%%%%%%%%%%%%%%%%%%%%%%%%%%%%%
We can use extrapolations from the Moseler data to calculate the resistivity, or equivalently the specular fraction $p$.  Substituting  \eqref{eq:rhoeff-linear} in \eqref{eq:FS}:
\begin{equation}
\label{eq:predict-p}
p = 1 - \frac{8}{3} \frac{L}{\lambda_b} \int_0^{2k_F} k_0 \: \zeta \left( k_0 \right) \; \frac{\tau_b}{l^2 \: \tau_s^{\rm eff}\left( k_0 \right)} \: dk_0,
\end{equation}
which (for $L>$100 nm) is independent of $L$.  
We find $p$ values of essentially 1 for the Gaussian PSD and $p=0.96$ ($1-p=0.04$) for \refeq{feenstra-fit}.  
We get the same results when we use the full PSDs directly as in Eqs.~(\ref{eq:W-def}) -- (\ref{eq:conductivity})%\footnote{Angular harmonics for gaussian and \refeq{feenstra-fit} given in \cite{MeyerPon}.}

\begin{figure}
  \includegraphics[scale=0.5]{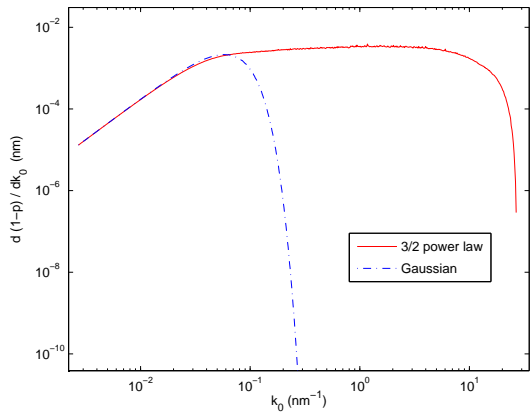}
  \caption{Integrand of \refeq{predict-p} with PSDs from Fig.~\ref{fig:exp-PSD}.  
Diffuse fraction $(1-p) \propto \rho_s^{\rm eff}$ is the area under the (1D) curve as shown.  The tail of the PSD is seen to have a major effect.  
The most important roughness components are $0.1 \: {\rm nm^{-1}} < k_0 < k_F = 13.6 \: {\rm nm^{-1}}$ for \refeq{feenstra-fit} and $k_0 \approx 0.05 \: {\rm nm^{-1}}$ for Gaussian. \label{fig:convolve} }
\end{figure}

%The overall resistivity from surface roughness is proportional the area under the curves in Fig.~\ref{fig:convolve}, as shown in Eqs.~\eqref{eq:rhoeff-linear} and \eqref{eq:predict-p}.  
The fact that our analysis predicts highly specular ($p \sim 1$) surface scattering for technologically-achievable surface roughness suggests that surface roughness scattering is a surmountable barrier to high conductivity. %in nanoscale wires. 
We can extract the most important components of roughness, taking into account both the relative strength of scattering and the observed roughness spectra.  %The fits to both \refeq{feenstra-fit} and Gaussian are extrapolations of the available experimental data beyond $k=0.1$ nm$^{-1}$.  Clearly the total resistivity depends strongly on the form of the PSD at higher wavenumbers.
The effective diffuse scattering rate as a function of spatial frequency is shown in Fig.~\ref{fig:convolve}. 
More accurate measurements of the high frequency portion of the PSD are clearly needed, as the frequencies above 0.1 nm$^{-1}$ are most critical to conductivity degradation.

Another way to understand these results is to note that for 
%a PSD of form 
\eqref{eq:feenstra-fit} with $\Lambda = 18$ nm, $p$ = 90\% corresponds to an RMS roughness of $7 \; \text{\AA}$, compared to experimental measurements in the range 2 -- 11 \AA \: \cite{RMS-refs, Purswani07}.  

The experimental literature is somewhat mixed on the relative importance of surface scattering.  Many results suggest that the observed resistivity increase is dominated by grain boundary rather than surface scattering \cite{GrainRefs}, while some extract values of $p$ near 0 (diffuse rather than specular scattering)~\cite{Purswani07}.  As we have seen, $p$ depends strongly ($O(l^2)$) on RMS roughness, which in turn depends on anneal times, deposition conditions, and other process variables.  Another interesting explanation for this discrepancy may come from the experiment of Rossnagel {\em et al.}~\cite{Rossnagel04}, who found that conductivity decreased strongly upon the deposition of an ultrathin Ta layer on top of a Cu film, but that the conductivity recovered when the Ta film was exposed to air, thereby oxidizing to become insulating.  These observations suggest that thin barrier/adhesion layers rather than surface/interface roughness may be causing the apparent diffuse surface scattering.

To summarize, we have found that bulk scattering can be included in quantum models of surface scattering by adding $\tau^{-1}$ matrices.  This leads to violation of Mathiessen's rule, but an  effective surface resistivity $\rho_s^{\rm{eff}}$ independent of bulk scattering can be extracted.  The resistivity from individual wavelengths of roughness can be convolved with roughness PSD to get $\rho_s^{\rm{eff}}$ for arbitrary surface.  %Extrapolating from experimental results gives modest diffuse scattering ($1-p \sim 0.04$).  
Our analysis suggests that roughness with wavelength within 1-2 orders of magnitude of the Fermi wavelength is the most critical for conductivity degradation.  

We would like to thank K.~Coffey and M.~Moseler for providing experimental details, and R.~Powell for useful discussions.  This work was supported by Intel Corporation and the Semiconductor Research Corporation (SRC).  

%%%%%%%%%%%%%%%%%%%%%%%%%%%%%%%%%%%%%%%%%%%%%%%%%%%%%%%%%%%%%%%%%%%%%%%%%%%%%%%%%%%%%%%%%%%%%%%%%%%%%%%%%%%%%%%%%%%%%%%%%%%%%%%%%%%

\end{document}